
%
%
%
%
\input phyzzx
%
%
\def\eqnon#1{\eqname{#1}}
\def\defeq{\equiv}
\def\={\defeq}

\def\sqr#1#2{{\vcenter{\hrule height.#2pt
             \hbox{\vrule width.#2pt height#1pt \kern#1pt
                   \vrule width.#2pt}
             \hrule height.#2pt}}}
\def\square{{\mathchoice{\sqr84}{\sqr84}{\sqr{5.0}3}{\sqr{3.5}3}}}

\def\rmand{{\rm and}}

\def\rmd{{\rm d}}

\def\rmEH{{\rm EH}}

\def\rmfor{{\rm for}}
\def\rmT{{\rm T}}

\def\rmtr{{\rm tr}}

\def\rmt{{\rm t}}
\def\rmL{{\rm L}}

\def\rms{{\rm s}}

\def\rmGF{{\rm GF}}

\def\rmFP{{\rm FP}}

\def\rmwith{{\rm with}}
\def\wt{\widetilde}

\mathchardef\bfalpha  ="090B
\mathchardef\bfbeta   ="090C
\mathchardef\bfgamma  ="090D
\mathchardef\bfdelta  ="090E
\mathchardef\bfepsilon="090F
\mathchardef\bfzeta   ="0910
\mathchardef\bfeta    ="0911
\mathchardef\bftheta  ="0912
\mathchardef\bfiota   ="0913
\mathchardef\bfkappa  ="0914
\mathchardef\bflambda ="0915
\mathchardef\bfmu     ="0916
\mathchardef\bfnu     ="0917
\mathchardef\bfxi     ="0918
\mathchardef\bfpi     ="0919
\mathchardef\bfrho    ="091A
\mathchardef\bfsigma  ="091B
\mathchardef\bftau    ="091C
\mathchardef\bfupsilon="091D
\mathchardef\bfphi    ="091E
\mathchardef\bfchi    ="091F
\mathchardef\bfpsi    ="0920
\mathchardef\bfomega  ="0921
\REF\HawZ{S.~W.~Hawking, Commun.~Math.~Phys.~{\bf 55},~133~(1977).}
\REF\GiPe{G.~W.~Gibbons and M.~J.~Perry,
Nucl.~Phys.~{\bf B146}, 90~(1978).}
\REF\HaHa{J.~B.~Hartle and S.~W.~Hawking,
Phys.~Rev.~{\bf D28},~2960~(1983).}
\REF\ChDu{S.~M.~Christensen and M.~J.~Duff,
Nucl.~Phys.~{\bf B170}[FS1], 480~(1980).}
\REF\Col{S.~Coleman, Nucl.~Phys.~{\bf B310}, 643~(1988).}
\REF\GHP{G.~W.~Gibbons, S.~W.~Hawking and M.~J.~Perry,
Nucl.~Phys.~{\bf B138},141~(1978).}
\REF\AbLe{E.~S.~Abers and B~.W.~Lee, Phys.~Rep.~{\bf 9},~1~(1973).}
\REF\Tei{C.~Teitelboim, Phys.~Rev.~{\bf D25},~3159~(1982).}
\REF\HaKu{J.~B.~Hartle and K.~V.~Kucha${\rm\check r}$,
in {\it Quantum Theory of Gravity},
edited by\break
S.~M.~Christensen
(Adam Hilger Ltd, Bristol, 1984).}
\REF\Hal{J.~J.~Halliwell, Phys.~Rev.~{\bf D38},~2468~(1988).}
\REF\HaSc{J.~B.~Hartle and K.~Schleich,
in {\it Quantum Field Theory and Quantum Statistics},
ed.~T.~A.~Batalin, C.~J.~Isham and G.~A.~Vilkovisky
(Hilger, Bristol, 1987):
K.~Schleich, Phys.~Rev.~{\bf D36}, 2342~(1987).}
\REF\AFKO{H.~Arisue, T.~Fujiwara, M.~Kato and K.~Ogawa,
Phys.~Rev.~{\bf D35}~(1987)~2309.}
\REF\FuKaL{H.~Fukutaka and T.~Kashiwa,
Prog.~Theor.~Phys.~{\bf 82}~(1989)~791.}
\REF\GrKo{P.~A.~Griffin and D.~A.~Kosower,
Phys. Lett. {\bf B233}, 295~(1989).}
\REF\Va{D.~V.~Vassilevich, Nuovo Cimento {\bf A104}, 743~(1991).}
\REF\FuF{H.~Fukutaka, Kyoto preprint~YITP/K-897~(1990) and
references therein.}
\REF\FuS{H.~Fukutaka, Kyoto preprint~YITP/K-958~(1992),
hepth@xxx/9201022.}
\REF\FaPo{L.~D.~Faddeev and V.~N.~Popov,
Phys.~Lett.~{\bf B25},~29~(1967).}
\REF\FaP{L.~D.~Faddeev, Theor.~Math.~Phys. {\bf 1}, 1~(1970).}
%
\noindent{\it Yukawa\hskip0.06cm Institute\hskip0.06cm Kyoto \hfill}
\vskip -1.0cm
\rightline{YITP/K-972}
\vskip -0.20cm
\rightline{hepth@xxx/9203049}
\vskip -0.20cm
\rightline{March 1992}

\title{Path-Integral Measure of
Linearized~Gravity in Curved Spacetime}

\centerline{{\twelverm H}{\tenrm IROKI\ }
{\twelverm \ F}{\tenrm UKUTAKA}}
\address{\null\hskip-8mm
Yukawa Institute for Theoretical Physics\break
Kyoto University,~Kyoto 606,~Japan}


\abstract{
The path-integral measure
of linearized gravity around a saddle-point background
with the cosmological term
is considered in order to study
the conformal rotation prescription
proposed by Gibbons, Hawking and Perry.
It is also argued that the most generally used measure, i.e.,
the covariant path-integral measure,
does not give us a one-loop partition function
which the only physical variables contribute
and that its path integral
fails to keep the cancellation of contributions
between the Faddeev-Popov ghosts
and the unphysical variables of the linearized gravitational field,
although it has a coordinate invariant measure.
In de~Sitter spacetime,
it is shown that the uncancellation factor can be understood as
a nontrivial (anomalous) Jacobian factor under the transformation
of the path-integral measure from covariant one to canonical one.}
\endpage

\chapter{Introduction}
The truth that Euclidean path integrals are directly combined
with classical actions
has made them most popular
as attractive and powerful tools for investigation of
field theories in curved spacetime
and quantum gravity [\HawZ-\Col].
They enable us to have suggestive discussions
with respect to gauge symmetries of systems
if the classical actions have them,
and also manifest coordinate invariance
which matches the spirit of general relativity.
Under such circumstances,
the issue of derivation
of the Euclidean gravitational path-integral formula
from the canonical quantization may be very important
to obtain theoretical meanings
of the conformal rotation prescription
proposed by Gibbons, Hawking and Perry [\GHP].
Furthermore, if it is done, objects given by the path integral
of gravity may become more meaningful,
for example, with respect to whether the path integral has
some information on the ground state
similarly to the case
of any field theory with a positive-definite hamiltonian
in the flat spacetime [\AbLe].

However, it is very difficult in the Einstein gravity
to derive the path integral or
to give the theoretical meanings to the conformal rotation.
Because the Einstein theory is a reparametrization one,
in which the physical time is a dynamical variable.
Thus it is necessary to identify the physical time
which plays a privileged role
in the canonical formalism,
when quantizing it according to that formalism [\Tei-\Hal].
With the identification
it might be possible to construct a path-integral formula
from the canonical approach.
Unfortunately, the choice of the physical time is a difficult problem
and even if it is done it is not so trivial
how the choice is reflected in the path-integral formula,
particularly in the covariant expression.
Hence the choice of the physical time may be an important issue
in quantum gravity.

Also in linearized gravity, we have to take the contour of integration
for a variable associated with the conformal factor
to be parallel to the imaginary axis
in order to make the Euclidean path integral converge,
since the action of linearized gravity is unbounded from below
with respect to the variable.
Here the time is not a dynamical variable,
so the theoretical foundation of the conformal rotation
in linearized gravity
may be different from that in the case of gravity,
but can be discussed from the standpoint as a gauge theory:
indeed, as studied by many authors
in the flat spacetime [\HaSc-\FuKaL],
the variable needing the conformal rotation is an unphysical one
like $A_0$ in the Abelian gauge theory,
and it is found that one has to rotate it to the imaginary axis,
in addition to Wick rotating,
in order to obtain the well-defined covariant formulation
of the Euclidean path integral.

Furthermore, in linearized gravity
around a nontrivial background such as
de~Sitter spacetime, it was shown by Griffin and Kosower [\GrKo]
that the Euclidean action of the physical part is positive definite,
which means that the conformal rotation is not necessary
when taking a special gauge fixing condition so that
redundant variables all may vanish
like the Coulomb gauge in the Abelian gauge theory,
although on this background the conformal rotation in covariant gauges
is more complicated as compared with that
on the flat spacetime background.
They also pointed out in this de~Sitter spacetime that
the covariant path integral does not give
the one-loop correction which is contributed
by the only physical variables,
in other words, that the covariant path integral is different
from the canonical path integral.
The same thing also happens
in the Abelian gauge theory in curved spacetime [\Va-\FuS].
This issue may be noticeable also from the point of view of
reexamination of the Faddeev-Popov conjecture [\FaPo,\FaP] that
path integrals in various gauge choices are all same,
because now the difference of the path-integral measures
may emerge as differences of path integrals
between various gauges [\FuS].

The purpose of this paper is to study
the relation between the covariant path integral
and the canonical approach
to linearized gravity in curved spacetime
and the uncancellation of the one-loop contributions
given from the unphysical variables and the Faddeev-Popov ghosts.
It is shown that the uncancellation factor can be also understood as
a nontrivial (anomalous) Jacobian factor under the transformation
of the path-integral measure from covariant one to canonical one,
similarly to the case of the Abelian gauge theory.

This paper is organized as follows:
in section 2
it is pointed out that the covariant path integral
of linearized gravity around a saddle-point background
with the cosmological term
does not give us the one-loop partition function
which the only physical modes contribute
and also that its path integral
fails to keep the cancellation of contributions
from the Faddeev-Popov ghosts
and the unphysical modes of the linearized gravitational field.
In section 3
we study how the modes needing the conformal rotation prescription
depend on the gauge fixing condition with gauge parameters $\alpha$
and $\eta$.
In section 4
it is shown that the uncancellation factor can be understood as
a nontrivial (anomalous) Jacobian factor under the transformation
of the path-integral measure from canonical one to covariant one.
Section 5 is devoted to conclusion,
and eigenfunctions on $S^D$ are summarized in Appendix A.

\chapter{BRST Path Integral of Linearized Gravity}

In this section, we formally calculate the one-loop correction
by the use of the BRST path-integral method with the covariant measure,
and a probability is then pointed out that
the contribution of the Faddeev-Popov ghosts
might not generally cancel out with
one from the unphysical degrees of freedom
of linearized gravitational field.

Our starting Euclidean path integral of linearized gravity
in a $D$-dimensional curved spacetime is
$$\eqalignno{
  {\cal Z}
&\=
  \int{\cal D}h_{\mu\nu}\Delta_{\rmFP}
  \exp\Big[ -{1\over\hbar}
  ( I+I^\rmGF )
  \Big],
&\eqnon\DefA\cr
}$$
where $I$ is the linearized gravity action
in Euclidean spacetime [\ChDu];
$$\eqalignno{
&
  I
 =
  {1\over4}\int\rmd^Dx
  \sqrt g\bigg( {1\over2}g^{\mu\nu}h-h^{\mu\nu} \bigg)
\cr
&\qquad
  \times\bigg(
  \square h_{\mu\nu}
  +\nabla_\mu\nabla_\nu h
  -2\nabla^\rho\nabla_\mu h_{\rho\nu}
  +{4\Lambda\over D-2}h_{\mu\nu}
  \bigg),
&\eqnon\KoLEz\cr
}$$
which is obtained by expanding the Einstein-Hilbert action
with the cosmological term
$$\eqalignno{
  I^\rmEH
&\=
  -{1\over\kappa^2}\int\rmd^Dx\sqrt{\bar g}
  (\bar R -2\Lambda ),
&\eqnon\IE\cr
}$$
around a classical background $g_{\mu\nu}$ which satisfies
$$\eqalignno{
&
  R_{\mu\nu}
  -{1\over2}Rg_{\mu\nu}
  +\Lambda g_{\mu\nu}
 =
  0,
&\eqnon\EqEin\cr
}$$
with
$
  \bar g_{\mu\nu}
 \=
  g_{\mu\nu}
  +\kappa h_{\mu\nu}
$,
and $I^\rmGF$ is a gauge fixing term
with gauge parameters $\alpha$ and $\eta$;
$$\eqalignno{
  I^\rmGF
&\=
  {1\over2\alpha}
  \int\rmd^Dx\sqrt g
  ( \nabla^\rho h_{\rho\mu}
  -\eta\nabla_\mu h
  )^2,
&\eqnon\DefIGF\cr
}$$
and then the Faddeev-Popov determinate, $\Delta_{\rmFP}$, becomes
$$\eqalignno{
  \Delta_{\rmFP}
&\=
  \Big\vert\det
  {1\over\sqrt\alpha}
  \Big[
  -\square\delta_\mu^\nu-\nabla^\nu\nabla_\mu
  +2\eta \nabla_\mu\nabla^\nu
  \Big](v)\Big\vert,
&\eqnon\FPdet\cr
}$$
where $v$ is any vector field.

Now, decompose $h_{\mu\nu}$ as
$$\eqalignno{
  h_{\mu\nu}
&=
  {1\over D}g_{\mu\nu}h
  +2\Big(
  \nabla_\mu\nabla_\nu\square^{-1}
  -{1\over D}g_{\mu\nu}
  \Big)
  \phi
  +\nabla_\mu\xi^\rmd_\nu
  +\nabla_\nu\xi^\rmd_\mu
  +h^{\wt{\rmtr}\rmd}_{\mu\nu},
&\eqnon\Defh\cr
}$$
where $h$ and $\phi$ are scalar fields,
$\xi_\mu^\rmd$ is a divergenceless vector field,
and $h^{\wt{\rmtr}\rmd}_{\mu\nu}$ is
a traceless and divergenceless tensor field:
$$\eqalign{
  \nabla^\mu\xi^\rmd_\mu
&=
  0,
\qquad
  g^{\mu\nu}h^{\wt{\rmtr}\rmd}_{\mu\nu}
 =
  \nabla^\rho h^{\wt{\rmtr}\rmd}_{\rho\mu}
 =
  0.
}\eqn\Decxi$$
Then, the actions $I$ and $I^\rmGF$ are separated
into four independent parts of their variables:
$$\eqalign{
&\hskip 3.0cm
  I
 =
  I^{\wt{\rmtr}\rmd}
  +I^{\psi},
\cr
  I^{\wt{\rmtr}\rmd}
&\=
  {1\over4}\int\rmd^Dx \sqrt g
  h^{\wt{\rmtr}\rmd}_{\mu\nu}
  \bigg[ \bigg(
  -\square
  +{4\Lambda\over(D-1)(D-2)}
  \bigg)g^{\mu\rho}g^{\nu\sigma}
  -2C^{\mu\rho\nu\sigma}
  \bigg]
  h^{\wt{\rmtr}\rmd}_{\rho\sigma},
\cr
  I^{\psi}
&\=
  {1\over4}\int\rmd^Dx \sqrt g
  \psi\Big( ( D-1 )( D-2 )\square
  +2D\Lambda
  \Big)\psi,
\cr
}\eqn\DecI
$$
and
$$\eqalign{
&\hskip 1.0cm
  I^{\rmGF}
 =
  I^{\rmGF^\rmd}
  +I^{\rmGF^\rms},
\cr
  I^{\rmGF^\rmd}
&\=
  {1\over2\alpha}
  \int\rmd^Dx\sqrt g
  g^{\mu\nu}\xi^\rmd_\mu
  \Big( \square +{2\Lambda\over D-2}
  \Big)^2\xi^\rmd_\nu,
\cr
  I^{\rmGF^\rms}
&\=
  -{2\over\alpha}
  \int\rmd^Dx\sqrt g
  \phi' \square^{-1}
  \Big( ( 1 -\eta )\square
  +{2\Lambda\over D-2}
  \Big)^2\phi',
\cr
}\eqn\GFs
$$
where $C_{\mu\rho\nu\sigma}$ is the Weyl tensor and
$$\eqalignno{
  \psi
&\=
  {1\over D}( h-2\phi ),
\qquad
  \phi'
 \=
  \phi
  -{\eta D-1\over2}\square
  \Big( (1-\eta)\square+{2\Lambda\over D-2} \Big)^{-1}\psi.
&\eqnon\Defphipra\cr
}$$
Obviously, $I^{\psi}$ may be negative definite
with respect to higher frequency modes than the mass $2D\Lambda$
owing to opposite sign of the kinetic term to ordinary scalar fields.
Thus, in addition to Wick rotating,
one has to always rotate the contour of integration
over these modes of $\psi$ to the imaginary axis
so as to make integrations converge.
But, it is not surprising because $\psi$ is an unphysical degree
of freedom of linearized gravitational field,
i.e., we can make the variable vanish as a consistency condition
for a suitable gauge fixing like the Coulomb gauge
whenever that is wanted,
although it is a gauge invariant variable.
Hence, if we start with the operator formalism,
the theoretical meaning of this rotation could be understood
similarly to the case in the flat spacetime [\HaSc-\FuKaL].
On the other hand, only $h^{\wt{\rmtr}\rmd}_{\mu\nu}$ has
the physical degrees of freedom,
which means that the path integral with respect to the physical mode
is well-defined within the positivity of $I^{\wt{\rmtr}\rmd}$,
which may be expected in any curved spacetime
as well as the flat and de~Sitter spacetimes.

Next, in order to carry on calculation of \DefA\
by the use of the covariant measure [\HawZ],
let us define the measure using the Polyakov measure
for convenience sake as
$$\eqalignno{
  \int{\cal D}h_{\mu\nu}
  \exp\Big[ -{1\over4\hbar}< h,h >\Big]
&=
  1,
&\eqnon\PIMh\cr
}$$
with
$$\eqalignno{
  < h,h >
&\=
  \int \rmd^Dx{1\over2}\sqrt g
  ( 2h^{\mu\nu}h_{\mu\nu} +Ch^2 ),
&\eqnon\DefPM\cr
}$$
in which the constant $C$ could be defined through derivation of
the path-integral formula from the canonical method,
hence here we may take $C=-1$.
Then we must integrate $h$ as an imaginary variable,
but this imaginary like integration does not always make
the Euclidean path integral \DefA\ completely converge.
(See section 3.)

Therefore \DefA\ becomes
$$\eqalignno{
  {\cal Z}
&=
  \Big\vert\det\Big[
  -g^{\mu\nu}
  \Big( \square+{2\Lambda\over D-2} \Big)\Big](v^\rmd) \Big\vert^{1/2}
  {\cal Z}^{\wt{\rmtr}\rmd},
&\eqnon\KoA\cr
}$$
where
$$\eqalignno{
  {\cal Z}^{\wt{\rmtr}\rmd}
&=
  \bigg\vert\det\bigg[ \bigg(
  -\square
  +{4\Lambda\over(D-1)(D-2)}
  \bigg)g^{\mu\rho}g^{\nu\sigma}
  -2C^{\mu\rho\nu\sigma}
  \bigg](t^{\wt{\rmtr}\rmd}) \bigg\vert^{-1/2}.
&\eqnon\KoAtrld\cr
}$$
Here $h_{\mu\nu}^{\wt\rmtr\rmd}$
includes the physical and unphysical modes,
hence,
only if the unphysical contributions in ${\cal Z}^{\wt\rmtr\rmd}$
cancel out with	the Faddeev-Popov ghost contribution, i.e.,
$$\eqalignno{
&
  \Big\vert\det\Big[
  -g^{\mu\nu}
  \Big( \square+{2\Lambda\over D-2} \Big)
  \Big](v^\rmd) \Big\vert^{1/2},
&\eqnon\FPOLC\cr
}$$
${\cal Z}$ gives just the physical contributions.
But it does not generally hold in curved spacetimes.
Indeed, in de~Sitter spacetime it happens,
as suggested by Griffin and Kosower [\GrKo].
Such an uncancellation arises also in the Abelian gauge theory
and it is shown that the uncancellation can be regarded
as the difference of measure between
the covariant and canonical path integrals [\FuF,\FuS].
We will study in section 4 the relation
between the uncancellation factor of unphysical contributions
in the covariant one-loop correction and
an anomalous Jacobian factor arising under the change of variables
in the Polyakov measure from covariant one to canonical one.

\chapter{Conformal Rotation}

In order to find out the modes needing the conformal rotation,
let us study the following eigenvalue equation;
$$\eqalignno{
  {2\over\sqrt g}
  {\delta\over\delta h_{\mu\nu}}
  \big( I+I^{\rmGF} \big)
&=
  \lambda h^{\mu\nu},
&\eqnon\EEQ\cr
}$$
where $\lambda$ is an eigenvalue and may be regarded as a constant.
Then we might find that
the modes with a negative eigenvalue,
whose contours of integration need to be rotated
to the imaginary axis in order to make integrations over them converge,
may change on values of positive $\alpha$ and real $\eta$,
but that the difference of number
between the positive and negative modes is,
of course, always unchanged,
which is obvious because of a simple reason that
the negative and positive parts of $I^\psi$ in \DecI\
does not change and the gauge fixing term in \DefIGF\ (\GFs)
is always positive independently of the gauge fixing parameters.

Eq.~\EEQ\ is separated into $h_{\mu\nu}^{\wt\rmtr\rmd}$,
$\xi_\mu^\rmd$ and scalar ($\psi$ and $\phi$) parts
by inserting \Defh:
$$\eqalignno{
  \bigg[ \bigg(
  -\square
  +{4\Lambda\over(D-1)(D-2)}
  \bigg)\delta_\mu^\rho\delta_\nu^\sigma
  -2C_{\mu\ \nu}^{\ \rho\ \sigma}
  \bigg]
  h^{\wt{\rmtr}\rmd}_{\rho\sigma}
&=
  \lambda h^{\wt{\rmtr}\rmd}_{\mu\nu},
&\eqnon\EEQh\cr
}$$
$$\eqalignno{
  \nabla_\mu
  \Big[
  -{1\over\alpha}
  \Big( \square +{2\Lambda\over D-2} \Big)
  \Big]
  \xi^\rmd_\nu
  +\nabla_\nu
  \Big[
  -{1\over\alpha}
  \Big( \square +{2\Lambda\over D-2} \Big)
  \Big]
  \xi^\rmd_\mu
&=
  \lambda (
  \nabla_\mu\xi_\nu^\rmd
  +\nabla_\nu\xi_\mu^\rmd
  ),
&\eqnon\EEQxi\cr
}$$
$$\eqalignno{
&\quad
  \Big[
  g_{\mu\nu}
  -\nabla_\mu\nabla_\nu
  \Big( \square +{2\Lambda\over D-2} \Big)^{-1}
  \Big]T^1
  +( \eta g_{\mu\nu} -\nabla_\mu\nabla_\nu\square^{-1} )T^2
\cr
&=
  \lambda
  ( g_{\mu\nu}\psi+2\nabla_\mu\nabla_\nu\square^{-1}\phi ),
&\eqnon\EEQS\cr
}$$
where
$$\eqalign{
  T^1
&\=
  (D-2)\Big( \square +{2\Lambda\over D-2} \Big)\psi,
\cr
  T^2
&\=
  -{2\over\alpha}
  \Big[
  (\eta D -1 )\square\psi
  +2\Big( (\eta -1)\square -{2\Lambda\over D-2} \Big)\phi
  \Big].
\cr
}\eqn\DefTab$$

Eq.~\EEQh\ says that if all eigenvalues of the operator
acting on traceless and divergenceless tensor fields, i.e.,
$$\eqalignno{
&
  \bigg[ \bigg(
  -\square
  +{4\Lambda\over(D-1)(D-2)}
  \bigg)\delta_\mu^\rho\delta_\nu^\sigma
  -2C_{\mu\ \nu}^{\ \rho\ \sigma}
  \bigg],
&\eqnon\Opeh\cr
}$$
are positive, we have no trouble when integrating out
$h^{\wt{\rmtr}\rmd}_{\mu\nu}$, that might be plausible
as mentioned in section 3.

{}From \EEQxi, eigenvalues of a tensor part defined by $\xi^\rmd_\mu$
are the same as eigenvalues of the operator
acting on divergence vector fields, i.e.,
$$\eqalignno{
&
  -{1\over\alpha}
  \Big( \square +{2\Lambda\over D-2} \Big)\delta^\rho_\mu.
&\eqnon\Opexid\cr
}$$
Here all eigenvalues of this operator are expected to be positive
because of the positivity of the gauge fixing action $I^\rmGF$,
so the Gaussian integrations
with respect to this part also converge.

To discuss about other remaining parts,
let us take, for simplicity,
$$\eqalignno{
  \alpha
&=
  {2(\eta D-1)\over D-2},
\cr
}$$
the trace part and the traceless part of $h_{\mu\nu}$
then are decoupled: Eq~\EEQS\ is reduced to two eigenvalue equations
$$\eqalign{
&\hskip 3.0cm
  {1\over D}g_{\mu\nu}
  \Big[
  (D-2)
  \Big( (1-\eta)\square+{2\Lambda\over D-2} \Big)
  \Big]h
 =
  \lambda{1\over D}g_{\mu\nu}h,
\cr
&
  \Big(
  \nabla_\mu\nabla_\nu\square^{-1}
  -{1\over D}g_{\mu\nu}
  \Big)
  \Big[
  -{2\over\alpha}
  \Big(
  (1-\eta)\square+{2\Lambda\over D-2}
  \Big)
  \Big]\phi
 =
  \lambda\Big(
  \nabla_\mu\nabla_\nu\square^{-1}
  -{1\over D}g_{\mu\nu}
  \Big)\phi.
\cr
}\eqn\EEQSS
$$
Thus eigenvalues of these tensor parts are the same as
these of the operators acting on scalar fields, such that
$$\eqalignno{
&
  (D-2)\Big( (1-\eta)\square+{2\Lambda\over D-2} \Big),
\qquad
  -{2\over\alpha}
  \Big(
  (1-\eta)\square+{2\Lambda\over D-2}
  \Big).
&\eqnon\EVS\cr
}$$
The $\eta$ dependence of these operators is very significant,
since the negative modes needing the conformal rotation
depend on its value and particularly for $\eta\ge1$ the variable
which must be rotated to
the imaginary axis is not the trace $h$ but $\phi$.
It may be also easily found from Eq.~\EVS\
that the difference of number
between the positive and negative modes of their operators
is unchanged as mentioned above,
because the signs of their operators are opposite to each other
always for all eigenvalues,
of course, although we must note here that
$\phi$ does not have some modes,
for example, the zero modes of $\square$, even if $h$ has the modes.
Furthermore, for the case of $\alpha=2(D-1)/(D-2)$ and $\eta=1$
we have an interesting gauge fixing in which
variables $h$ and $\phi$ have no kinetic terms but only mass terms,
contrary to the case of the flat spacetime background.

Finally, we may note that Eq.~\KoA\ can be directly checked
by Eqs.~\EEQh, \EEQxi\ and \EEQSS.

\chapter{Anomalous Jacobian Factor}

In this section, we take the Euclidean de~Sitter background
and discuss a relation between
the uncanceled factor
in the one-loop contributions given
by the covariant path integral of linearized gravity
and the nontrivial Jacobian factor under the transformation from
the covariant measure to the canonical measure,
choosing coordinates so that the metric has the form
$$\eqalign{
  \rmd s^2
&=
  \rmd \tau^2 +a^2(\tau)\rmd\Omega_{D-1}^{\ 2},
\qquad
  \rmd\Omega_{D-1}^{\ 2}
 =
  {\wt g}_{ij}\rmd x^i\rmd x^j,
\cr
  x^D
&=
  \tau
 =
  r\theta,
\qquad
  a
 =
  r\sin(\tau/r),
\qquad
  r^{-2}
 \=
  {2\Lambda\over(D-1)(D-2)}.
\cr
}\eqn\Defr
$$
On this manifold, we have $C_{\mu\nu\rho\sigma}=0$,
so the uncancellation factor we have to study
becomes, from \KoA\ together with formulae in Appendix A
(or ref.~[\GrKo]), as
$$\eqalignno{
&
  \Big\vert\det\Big[ g^{\mu\nu}
  \Big( -\square-(D-1)r^{-2} \Big)\Big](v^\rmd) \Big\vert^{1/2}
  \Big\vert\det\Big[
  g^{\mu\nu}
  \Big( -\square +r^{-2} \Big)\Big](\wt v^\rmd) \Big\vert^{-1/2},
&\eqnon\UCFc\cr
}$$
where $\wt v^\rmT$ is any divergenceless vector field
with no $\ell=1$ modes of
the covariant Laplacian $\wt\square$ on the unit $S^{D-1}$.
\vskip 0.5cm
{\bf 4.1\quad Covariant and Canonical Path-Integral Measures}

In this subsection, we construct the Polyakov measure
for tensor fields on $S^D$ by using two decompositions, i.e.,
the covariant and canonical decompositions.

The covariant decomposition in Eq.~\Defh\ may be rewritten as
$$\eqalignno{
  h_{\mu\nu}
&\=
  h^\rmtr_{\mu\nu}
  +h^{\wt\rmtr\phi}_{\mu\nu}
  +h^{\wt\rmtr\xi^\rmd}_{\mu\nu}
  +h^{\wt\rmtr\rmd}_{\mu\nu},
&\eqnon\DefhS\cr
}$$
where components $h^{\rmtr}_{\mu\nu}$ and $h^{\wt\rmtr\phi}_{\mu\nu}$
are defined by
$$\eqalignno{
  h^{\rmtr}_{\mu\nu}
&\=
  {1\over D}g_{\mu\nu}h
 \=
  {1\over\sqrt D}g_{\mu\nu}\bar h,
&\eqnon\Defhtr\cr
  h^{\wt\rmtr\phi}_{\mu\nu}
&\=
  2\Big(
  \nabla_\mu\nabla_\nu\square^{-1}
  -{1\over D}g_{\mu\nu}
  \Big)\phi
 \=
  \Big(
  \nabla_\mu\nabla_\nu
  -{1\over D}g_{\mu\nu}\square
  \Big)O_\phi^s\bar\phi,
&\eqnon\Defhwtrp\cr
  O_\phi^s
&\=
  \bigg(
  {D\over(D-1)(-\square)(-\square-Dr^{-2})}
  \bigg)^{1/2},
&\eqnon\DefOphi\cr
}$$
and $h^{\wt\rmtr\xi^\rmd}_{\mu\nu}$ is
$$\eqalignno{
  h^{\wt\rmtr\xi^\rmd}_{\mu\nu}
&\=
  \nabla_\mu\xi^\rmd_\nu
  +\nabla_\nu\xi^\rmd_\mu
 \=
  \nabla_\mu O_{\xi^\rmd}^v\bar\xi^\rmd_\nu
  +\nabla_\nu O_{\xi^\rmd}^v\bar\xi^\rmd_\mu,
&\eqnon\DefhwtrxdIJ\cr
  O_{\xi^\rmd}^v
&\=
  \Big( 2\big( -\square-(D-1)r^{-2} \big) \Big)^{-1/2},
&\eqnon\DefOxd\cr
}$$
which can be separated, according to whether
$\xi^\rmd_\mu$ is a longitudinal vector $\xi^\rmL_\mu$
or a transverse vector $\xi^\rmT_\mu$,
into $h^{\wt\rmtr\xi^\rmL}_{\mu\nu}$ and
$h^{\wt\rmtr\xi^\rmT}_{\mu\nu}$:
$$\eqalignno{
  h^{\wt\rmtr\xi^\rmd}_{\mu\nu}
&\=
  h^{\wt\rmtr\xi^\rmL}_{\mu\nu}
  +h^{\wt\rmtr\xi^\rmT}_{\mu\nu},
&\eqnon\DefphwtrxdIJ\cr
}$$
where
$$\eqalign{
  \bar\xi^\rmd_\mu
&\=
  \bar\xi^\rmL_\mu+\bar\xi^\rmT_\mu,
\qquad
  \bar\xi^\rmT_D
 =
  \wt\nabla^\ell
  \bar\xi^\rmT_\ell
 =
  0,
\cr
  \bar\xi^\rmL_D
&\=
  a^{-1}
  O_{\xi^\rmL}^s
  \bar\xi,
\qquad
  \bar\xi^\rmL_i
 \=
  -\wt\nabla_i\wt\square^{-1}
  a^{-(D-3)}\partial_D\big( a^{D-1}\bar\xi^\rmL_D \big),
\cr
  O_{\xi^\rmL}^s
&\=
  \bigg[
  {-\wt\square\over -\square+(D-2)r^{-2}}
  \bigg]^{1/2},
\cr
}\eqn\Defxid
$$
with the covariant derivative $\wt\nabla_i$ and
the covariant Laplacian $\wt\square$ on the unit $S^{D-1}$.
The last component $h^{\wt\rmtr\rmd}_{\mu\nu}$ in Eq.~\DefhS\
is decomposed into three parts:
$$\eqalignno{
  h^{\wt\rmtr\rmd}_{\mu\nu}
&\=
  h^{\wt\rmtr\rmd\rmL}_{\ \mu\nu}
  +h^{\wt\rmtr\rmd\rmT}_{\ \mu\nu}
  +h^{\rmt\rmT}_{\mu\nu},
&\eqnon\hwtrd\cr
}$$
where each part, in the coordinate system \Defr, is defined by
$$\eqalign{
  h^{\wt\rmtr\rmd\rmL}_{DD}
&\=
  h^{\wt\rmtr\rmd}_{DD}
 \=
  a^{-2}O_{\wt\rmtr\rmd\rmL}^s
  \bar h^{\wt\rmtr\rmd\rmL}_{DD},
\cr
  h^{\wt\rmtr\rmd\rmL}_{\ iD}
&\=
  \wt\nabla_i\wt\square^{-1}
  \wt\nabla^\ell h^{\wt\rmtr\rmd}_{\ell D}
 =
  -\wt\nabla_i\wt\square^{-1}
  a^{-(D-2)}\partial_D
  \big(
  a^Dh^{\wt\rmtr\rmd\rmL}_{DD}
  \big),
\cr
  h^{\wt\rmtr\rmd\rmL}_{\ ij}
&\=
  {D-1\over D-2}
  \Big(
  \wt\nabla_i\wt\nabla_j\wt\square^{-1}-{1\over D-1}\wt g_{ij}
  \Big)
  ( \wt\square+D-1 )^{-1}
\cr
&\qquad
  \times
  \Big[
  a^{-(D-3)}\partial_D
  \big( a\partial_D( a^Dh^{\wt\rmtr\rmd\rmL}_{DD} )\big)
  +{a^2\over D-1}\wt\square h^{\wt\rmtr\rmd\rmL}_{DD}
  \Big]
  -{a^2\over D-1}\wt g_{ij}h^{\wt\rmtr\rmd\rmL}_{DD},
\cr
  O_{\wt\rmtr\rmd\rmL}^s
&\=
  \bigg[
  { (D-2)(-\wt\square)\big( -\wt\square-(D-1) \big)
  \over(D-1)(-\square)\big( -\square+(D-2)r^{-2} \big) }
  \bigg]^{1/2},
\cr
}\eqn\htldL
$$
$$\eqalign{
  h^{\wt\rmtr\rmd\rmT}_{DD}
&\=
  \bar h^{\wt\rmtr\rmd\rmT}_{DD}
 \=
  0,
\qquad
  h^{\wt\rmtr\rmd\rmT}_{\ iD}
 \=
  a^{-1}O_{\wt\rmtr\rmd\rmT}^v
  \bar h^{\wt\rmtr\rmd\rmT}_{\ iD}
 \=
  h^{\wt\rmtr\rmd}_{iD}
  -\wt\nabla_i\wt\square^{-1}
  \wt\nabla^\ell h^{\wt\rmtr\rmd}_{\ell D},
\cr
  h^{\wt\rmtr\rmd\rmT}_{\ ij}
&\=
  -\wt\nabla_i( \wt\square+D-2 )^{-1}
  a^{-(D-3)}\partial_D( a^{D-1}h^{\wt\rmtr\rmd\rmT}_{\ jD} )
\cr
&\quad
  -\wt\nabla_j( \wt\square+D-2 )^{-1}
  a^{-(D-3)}\partial_D( a^{D-1}h^{\wt\rmtr\rmd\rmT}_{\ iD} ),
\cr
  O_{\wt\rmtr\rmd\rmT}^v
&\=
  \bigg[
  { -\wt\square-(D-2)\over2(-\square+r^{-2}) }
  \bigg]^{1/2},
\cr
}\eqn\htldT
$$
and
$$\eqalign{
  h^{\rmt\rmT}_{\mu D}
&\=
  0,
\qquad
  \wt g^{ij}h^{\rmt\rmT}_{ij}
 \=
  0,
\qquad
  \wt\nabla^\ell
  h^{\rmt\rmT}_{\ell i}
 \=
  0.
\cr
}\eqn\htT
$$
Then Eq.~\DefPM\ becomes
$$\eqalignno{
  < h,h' >
&=
  < h^\rmtr,h'^\rmtr >
  +< h^{\wt\rmtr\phi},h'^{\wt\rmtr\phi} >
  +< h^{\wt\rmtr\xi^\rmL},h'^{\wt\rmtr\xi^\rmL} >
  +< h^{\wt\rmtr\xi^\rmT},h'^{\wt\rmtr\xi^\rmT} >
\cr
&\quad
  +< h^{\wt\rmtr\rmd\rmL},h'^{\wt\rmtr\rmd\rmL} >
  +< h^{\wt\rmtr\rmd\rmT},h'^{\wt\rmtr\rmd\rmT} >
  +< h^{\rmt\rmT},h'^{\rmt\rmT} >,
&\eqnon\KeiSIPhtrl\cr
}$$
with
$$\eqalign{
&
  < h^\rmtr,h'^\rmtr >
 =
  \int\rmd^Dx\sqrt g
  {CD+2\over2}\bar h\bar h',
\qquad
  < h^{\wt\rmtr\phi},h'^{\wt\rmtr\phi} >
 =
  \int\rmd^Dx\sqrt g
  \bar\phi
  \bar\phi',
\cr
&
  < h^{\wt\rmtr\xi^\rmL},h'^{\wt\rmtr\xi^\rmL} >
 =
  \int\rmd^Dx\sqrt g
  \bar\xi\bar\xi',
\qquad\quad
  < h^{\wt\rmtr\rmd\rmL},h'^{\wt\rmtr\rmd\rmL} >
 =
  \int\rmd^Dx\sqrt g
  \bar h^{\wt\rmtr\rmd\rmL}_{DD}
  \bar h'^{\wt\rmtr\rmd\rmL}_{DD},
\cr
&\hskip 3.0cm
  < h^{\wt\rmtr\xi^\rmT},h'^{\wt\rmtr\xi^\rmT} >
 =
  \int\rmd^Dx\sqrt g
  g^{\mu\nu}\bar\xi^\rmT_\mu
  \bar\xi'^\rmT_\nu,
\cr
&\hskip 3.0cm
  < h^{\wt\rmtr\rmd\rmT},h'^{\wt\rmtr\rmd\rmT} >
 =
  \int\rmd^Dx\sqrt gg^{\mu\nu}
  \bar h^{\wt\rmtr\rmd\rmT}_{\mu D}
  \bar h'^{\wt\rmtr\rmd\rmT}_{\nu D},
\cr
&\hskip 3.0cm
  < h^{\rmt\rmT},h'^{\rmt\rmT} >
 =
  \int\rmd^Dx\sqrt gg^{\mu\rho}g^{\nu\sigma}
  h^{\rmt\rmT}_{\mu\nu}
  h'^{\rmt\rmT}_{\rho\sigma}.
\cr
}\eqn\CDTP
$$
Therefore, the Gaussian integral, i.e.,
$
  \int^{+\infty}_{-\infty}\rmd x
  e^{-\lambda x^2}
 =
  \sqrt{\pi/\lambda}
$,
says that the Polyakov measure \PIMh\ is represented in terms of
$\bar h$, $\bar\phi$, $\bar\xi$, $\bar h^{\rmt\rmd\rmL}_{DD}$,
$\bar\xi^\rmT_\mu$, $\bar h^{\wt\rmtr\rmd\rmT}_{\mu D}$ and
$h^{\rmt\rmT}_{\mu\nu}$ as
$$\eqalign{
&
  \int
  {{\cal D}\bar h\over\sqrt{8\pi\hbar/(CD+2)}}
  {{\cal D}\bar\phi\over\sqrt{4\pi\hbar}}
  {{\cal D}\bar\xi\over\sqrt{4\pi\hbar}}
  {{\cal D}\bar h^{\rmt\rmd\rmL}_{DD}\over\sqrt{4\pi\hbar}}
\cr
&\hskip -0.6cm
  \times
  {{\cal D}\bar\xi^\rmT_\mu\over\sqrt{4\pi\hbar}}
  {{\cal D}\bar h^{\wt\rmtr\rmd\rmT}_{\mu D}\over\sqrt{4\pi\hbar}}
  {{\cal D}h^{\rmt\rmT}_{\mu\nu}\over\sqrt{4\pi\hbar}}
  \exp\Big[ -{1\over4\hbar}< h,h > \Big]
 =
  1,
\cr
}\eqn\PMCD
$$
where integration variables may be concretely defined in terms of
the expansion coefficients of each variable
in the basis of the eigenfunctions of $\square$, i.e.,
those of associated components of $h_{\mu\nu}$
in the basis of the tensor eigenfunctions of $\square$
on $S^D$. (As for eigenfunctions on $S^D$ see Appendix A.)

In the canonical decomposition, which is most suitable one
for the case of a special gauge in which all redundant variables vanish
like the Coulomb gauge in the Abelian gauge theory
and also for a path integral in the phase space
where the time coordinate is separated from the other coordinates,
any tensor field can be expressed with
$$\eqalignno{
  h_{\mu\nu}
&\=
  h^\rmtr_{\mu\nu}
  +h^{h^{\wt\rmtr}_{DD}}_{\mu\nu}
  +h^{\Phi}_{\mu\nu}
  +h^{\psi}_{\mu\nu}
  +h^{h^\rmT}_{\mu\nu}
  +h^{w^\rmT}_{\mu\nu}
  +h^{\rmt\rmT}_{\mu\nu},
&\eqnon\NDefhIJ\cr
}$$
where
$$\eqalign{
  h^{h^{\wt\rmtr}_{DD}}_{DD}
&\=
  h^{\wt\rmtr}_{DD},
\qquad
  h^{h^{\wt\rmtr}_{DD}}_{iD}
 \=
  0,
\qquad
  h^{h^{\wt\rmtr}_{DD}}_{ij}
 \=
  -{1\over D-1}\wt g_{ij}a^2h^{\wt\rmtr}_{DD},
\cr
  h^\Phi_{iD}
&\=
  \wt\nabla_i\wt\square^{-1}\Phi,
\qquad
  h^\Phi_{DD}
 \=
  h^\Phi_{ij}
 \=
  0,
\cr
  h^{\psi}_{ij}
&\=
  {D-1\over D-2}
  \Big(
  \wt\nabla_i\wt\nabla_j\wt\square^{-1}
  -{1\over D-1}\wt g_{ij}
  \Big)( \wt\square+D-1 )^{-1}\psi,
\quad
  h^{\psi}_{\mu D}
 \=
  0,
\cr
  h^{h^\rmT}_{iD}
&\=
  h^\rmT_{iD},
\qquad
  h^{h^\rmT}_{DD}
 \=
  h^{h^\rmT}_{ij}
 \=
  0,
\qquad
  \wt\nabla^\ell h^\rmT_{\ell D}
 \=
  0,
\cr
  h^{w^\rmT}_{ij}
&\=
  \wt\nabla_i( \wt\square+D-2 )^{-1}w^\rmT_j
  +\wt\nabla_j( \wt\square+D-2 )^{-1}w^\rmT_i,
\qquad
  h^{w^\rmT}_{\mu D}
 \=
  0,
\qquad
  \wt\nabla^\ell w^{\rmT}_\ell
 \=
  0.
\cr
}\eqn\CAD
$$
Then Eq.~\DefPM\ is split into each part of them:
$$\eqalignno{
  < h,h' >
&=
  < h^\rmtr,h'^\rmtr >
  +< h^{h^{\wt\rmtr}_{DD}},h'^{h^{\wt\rmtr}_{DD}} >
  +< h^{\Phi},h'^{\Phi} >
  +< h^{\psi},h'^{\psi} >
\cr
&\quad
  +< h^{h^\rmT},h'^{h^\rmT} >
  +< h^{w^\rmT},h'^{w^\rmT} >
  +< h^{\rmt\rmT},h'^{\rmt\rmT} >,
&\eqnon\KeiSIPhtrl\cr
}$$
with
$$\eqalign{
  < h^\rmtr,h'^\rmtr >
&=
  \int\rmd\tau {CD+2\over2D}a^{D-1}
  \int\rmd^{D-1}x\sqrt{\wt g}
  hh',
\cr
  < h^{h^{\wt\rmtr}_{DD}},h'^{h^{\wt\rmtr}_{DD}} >
&=
  \int\rmd\tau
  {D\over D-1}
  a^{D-1}
  \int\rmd^{D-1}x\sqrt{\wt g}
  h^{\wt\rmtr}_{DD}h'^{\wt\rmtr}_{DD},
\cr
  < h^{\Phi},h'^{\Phi} >
&=
  \int\rmd\tau2a^{D-3}
  \int\rmd^{D-1}x\sqrt{\wt g}
  \Phi(-\wt\square^{-1})\Phi',
\cr
  < h^\psi,h'^\psi >
&=
  \int\rmd\tau a^{D-5}
  \int\rmd^{D-1}x\sqrt{\wt g}
  {D-1\over D-2}
  \psi
  ( \wt\square+D-1 )^{-1}
  \wt\square^{-1}
  \psi',
\cr
 < h^{h^\rmT},h'^{h^\rmT} >
&=
  \int\rmd\tau 2a^{D-3}
  \int\rmd^{D-1}x\sqrt{\wt g}\wt g^{ij}
  h^\rmT_{iD}h'^\rmT_{jD},
\cr
  < h^{w^\rmT},h'^{w^\rmT} >
&=
  \int\rmd\tau 2a^{D-5}
  \int\rmd^{D-1}x\sqrt{\wt g}\wt g^{ij}
  w^\rmT_i
  \big( -\wt\square-(D-2) \big)^{-1}
  w'^\rmT_j,
\cr
  < h^{\rmt\rmT},h'^{\rmt\rmT} >
&=
  \int\rmd\tau a^{D-5}
  \int\rmd^{D-1}x\sqrt{\wt g}
  \wt g^{im}\wt g^{jn}
  h^{\rmt\rmT}_{ij}h'^{\rmt\rmT}_{mn}.
\cr
}\eqn\CADTP
$$
Therefore, the Gaussian integral says that
the Polyakov measure \PIMh\ in the canonical decomposition
becomes
$$\eqalign{
&
  \int\prod_\tau
  \Big[
  \Big( {\Delta\tau(CD+2)a^{D-1}(\tau)\over8\pi\hbar D} \Big)^{1/2}
  {\cal D}h(\tau)
  \Big]
  \Big[
  \Big( {\Delta\tau Da^{D-1}(\tau)\over4\pi\hbar(D-1)} \Big)^{1/2}
  {\cal D}h^{\wt\rmtr}_{DD}(\tau)
  \Big]
\cr
&\quad
  \times
  \Big[
  \Big( {\Delta\tau a^{D-3}(\tau)
  \over2\pi\hbar(-\wt\square) } \Big)^{1/2}
  {\cal D}\Phi(\tau)
  \Big]
  \Big[
  \Big(
   {\Delta\tau(D-1)a^{D-5}(\tau)\over4\pi\hbar(D-2)
  ( \wt\square+D-1 )\wt\square
  } \Big)^{1/2}
  {\cal D}\psi(\tau)
  \Big]
\cr
&\quad
  \times
  \Big[
  \Big( {\Delta\tau a^{D-3}(\tau)\over2\pi\hbar} \Big)^{1/2}
  {\cal D}h^\rmT_{iD}(\tau)
  \Big]
  \Big[
  \Big(
  {\Delta\tau a^{D-5}(\tau)\over2\pi\hbar
  \big( -\wt\square-(D-2) \big) }
  \Big)^{1/2}
  {\cal D}w^\rmT_i(\tau)
  \Big]
\cr
&\quad
  \times
  \Big[
  \Big( {\Delta\tau a^{D-5}(\tau)\over4\pi\hbar} \Big)^{1/2}
  {\cal D}h^{\rmt\rmT}_{ij}(\tau)
  \Big]
  \exp\Big[
  -{1\over4\hbar}
  < h,h >
  \Big]
 =
  1,
\cr
}\eqn\PMCAD
$$
in which the time $\tau$ is specialized from the other coordinates,
and its product might be defined in the discrete time formulation
with a finite distance and its zero limitation
after integrating out,
and then the functional measure on each time are defined
in terms of the expansion coefficients of each variable of
$h$, $h^{\wt\rmtr}_{DD}$, $\Phi$, $\psi$, $h^\rmT_{iD}$,
$w^\rmT_i$ and ${\cal D}h^{\rmt\rmT}_{ij}$
in the basis of the eigenfunctions of $\wt\square$
on the unit $S^{D-1}$.

Eqs.~\PMCD\ and \PMCAD\ mean that
the Jacobian factor under the transformation of measure
from the covariant path integral to the canonical one
would be naively 1,
since both measures are defined with the Gaussian integral
whose integration value is 1.
However, as discussed in the following subsections,
the Jacobian factor is unfortunately not 1 and takes an anomalous value.
Such a situation happens also in the case of
the gauge theory in Euclidean Robertson-Walker spacetimes
with $K=+1$ [\FuF,\FuS].
The Jacobian factor is discussed below,
separating each one of two decompositions
into three parts, i.e.,
a traceless and transverse tensor part,
a tensor part defined by transverse vector fields
$\bar\xi^\rmT_i$, $\bar h^{\wt\rmtr\rmd\rmT}_{\ iD}$,
$h^\rmT_{iD}$ and $w^\rmT_i$,
and that defined by scalar fields
$\bar h$, $\bar\phi$, $\bar\xi$, $\bar h^{\wt\rmtr\rmd\rmL}_{DD}$,
$h$, $h^{\wt\rmtr}_{DD}$, $\Phi$ and $\psi$.

As for the physical variable, i.e., the traceless and transverse
tensor $h^{\wt\rmtr\rmT}_{\mu\nu}$, it is easily found that
we have no anomalous thing under the transformation between
the covariant and canonical measures:
$$\eqalignno{
  \prod_\tau\Big[
  \Big( {\Delta\tau a^{D-5}(\tau)\over4\pi\hbar} \Big)^{1/2}
  {\cal D}h^{\rmt\rmT}_{ij}(\tau)
  \Big]
&=
  {{\cal D}h^{\rmt\rmT}_{\mu\nu}\over\sqrt{4\pi\hbar}},
&\eqnon\AJPHYP\cr
}$$
where it is noted that the time $\tau$ product
in the l.h.s.~is changed, in the r.h.s., into the product
with respect to modes along the time axis,
by using the relation (A32) among
the traceless and transverse tensor eigenfunctions on $S^D$
and the traceless and divergenceless eigenfunctions on $S^{D-1}$,
and then the factors
$\prod_\tau\big( \Delta\tau a^{D-5}(\tau) \big)^{1/2}$
cancel out with the determinants of $a^2f^{L\ell}$, since
$$\eqalignno{
  \vert\det f^{L\ell}\vert
&=
  \prod_\tau\big( \Delta\tau a^{D-1}(\tau) \big)^{-1/2}.
&\eqnon\detfLell\cr
}$$
(See Ref.~\FuF.)
\vskip 0.5cm
{\bf 4.2\quad Anomalous Jacobian Factor in Transverse Vector Part}

The relations among the transverse vector fields
$\bar\xi^\rmT_i$, $\bar h^{\wt\rmtr\rmd\rmT}_{\ iD}$,
$h^\rmT_{iD}$ and $w^\rmT_i$ used in
the covariant and canonical decompositions,
i.e., \DefhS\ and \NDefhIJ, are
$$\eqalignno{
&
  h^\rmT_{iD}
 =
  a^2\partial_D\big( a^{-2}O_{\xi^\rmd}^v\bar\xi^\rmT_i \big)
  +a^{-1}O_{\wt\rmtr\rmd\rmT}^v\bar h^{\wt\rmtr\rmd\rmT}_{\ iD},
&\eqnon\KanhTiD\cr
&
  (\wt\square+D-2)^{-1}w^\rmT_i
 =
  O_{\xi^\rmd}^v\bar\xi^\rmT_i
  -(\wt\square+D-2)^{-1}a^{-(D-3)}
  \partial_D
  \big(
  a^{D-2}O_{\wt\rmtr\rmd\rmT}^v\bar h^{\wt\rmtr\rmd\rmT}_{\ iD} \big),
&\eqnon\KanwTi\cr
&
  a^2\big( \square+(D-1)r^{-2} \big)O_{\xi^\rmd}^v\bar\xi^\rmT_i
 =
  a^{-(D-3)}\partial_D\big( a^{D-1}h^\rmT_{iD} \big)
  +w^\rmT_i,
&\eqnon\KanxiTi\cr
&
  a(\wt\square+D-2)^{-1}
  \big(
  \square
  -r^{-2}
  \big)
  O_{\wt\rmtr\rmd\rmT}^v\bar h^{\wt\rmtr\rmd\rmT}_{\ iD}
 =
  h^\rmT_{iD}
  -a^2\partial_D
  \big( a^{-2}
  (\wt\square+D-2)^{-1}w^\rmT_i
  \big),
&\eqnon\KanhtrdTiD\cr
}$$
where the relation between the former two equations
and the later two equations is that of the inverse transformation.
Now let us discuss an anomalous Jacobian factor
with respect to these variables, separating them
into $\ell=1$ modes and $\ell\neq1$ modes of $\wt\square$.

First, with respect to $\ell=1$ modes,
noting that $h^{\rmT}_{iD}$ and $\bar\xi^{\rmT}_i$ have their modes,
on the other hand,
$\bar h^{\wt\rmtr\rmd\rmT}_{\ iD}$ and $w^\rmT_i$ are not able
to have them, the following relations
between $h^{\rmT}_{iD}$ and $\bar\xi^{\rmT}_i$ are derived
from Eqs.~\KanhTiD\ and
\KanxiTi\ (although both Eqs.~\KanwTi\ and \KanhtrdTiD\ are reduced
to a trivial equation);
$$\eqalignno{
&\quad
  h^{\rmT(\wt1)}_{\ iD}
 =
  a^2\partial_D
  \big( a^{-2}O_{\xi^\rmd}^v\bar\xi^{\rmT(\wt1)}_{\ \ i} \big),
&\eqnon\KanhTiDa\cr
&\quad
  \big( \square+(D-1)r^{-2} \big)
  O_{\xi^\rmd}^v\bar\xi^{\rmT(\wt1)}_{\ \ i}
 =
  a^{-(D-1)}
  \partial_D
  \big( a^{D-1}h^{\rmT(\wt1)}_{\ iD} \big),
&\eqnon\KanxiTia\cr
}$$
where the indices $(\wt1)$ means that the variables are
of $\ell=1$ modes and these equations have the inverse relationship
to each other.
{}From \KanhTiDa, a nontrivial Jacobian factor is obtained:
$$\eqalignno{
&
  \Big\vert\det\Big[ a^2\partial_Da^{-2} \Big](v^{\rmT(\wt1)}) \Big\vert
  \Big\vert\det\Big[
  -g^{\mu\nu}\big( \square+(D-1)r^{-2} \big) \Big](v^{\rmT(\wt1)})
  \Big\vert^{-1/2},
&\eqnon\tvaAJA\cr
}$$
or if Eq.~\KanxiTia\ is used to change the variables,
another Jacobian factor is
$$\eqalignno{
&
  \Big\vert\det\Big[
  a^{-(D-1)}\partial_Da^{D-1} \Big](v^{\rmT(\wt1)})
  \Big\vert^{-1}
  \Big\vert\det\Big[
  -g^{\mu\nu}\big( \square+(D-1)r^{-2} \big) \Big](v^{\rmT(\wt1)})
  \Big\vert^{1/2},
&\eqnon\tvbAJA\cr
}$$
where $v^{\rmT(\wt1)}$ is any transverse vector field with
$\ell=1$ modes of $\wt\square$ on the unit $S^{D-1}$.

With respect to $\ell\ne1$ part,
if Eqs.~\KanxiTi\ and \KanhTiD\ are used under the changes of
variables,
$
  \big( h^\rmT_{iD},w^\rmT_i \big)
 \rightarrow
  \big( h^\rmT_{iD},\bar\xi^\rmT_\mu \big)
 \rightarrow
  \big( \bar h^{\wt\rmtr\rmd\rmT}_{\ \mu D},\bar\xi^\rmT_\mu \big)
$,
then we have
$$\eqalignno{
&\quad
  \prod_\tau
  \Big[
  \Big(
  {\Delta\tau a^{D-5}(\tau)\over2\pi\hbar
  \big( -\wt\square-(D-2) \big) }
  \Big)^{1/2}
  {\cal D}w^\rmT_i(\tau)
  \Big]
\cr
&=
  \prod_\tau \big( a(\tau) \big)
  \bigg\vert\det\bigg[
  g^{\mu\nu}\bigg(
  { -\square-(D-1)r^{-2} \over -\wt\square-(D-2) }
  \bigg) \bigg](\wt v^\rmT)
  \bigg\vert^{1/2}
  {{\cal D}\bar\xi^\rmT_\mu\over\sqrt{4\pi\hbar}},
&\eqnon\FCVa\cr
}$$
and
$$\eqalignno{
&\quad
  \prod_\tau
  \Big[
  \Big( {\Delta\tau a^{D-3}(\tau)\over2\pi\hbar} \Big)^{1/2}
  {\cal D}h^\rmT_{iD}(\tau)
  \Big]
\cr
&=
  \prod_\tau\big( a^{-1}(\tau) \big)
  \bigg\vert\det\bigg[
  g^{\mu\nu}\bigg(
  { -\wt\square-(D-2) \over -\square+r^{-2} }
  \bigg) \bigg](\wt v^\rmT)
  \bigg\vert^{1/2}
  {{\cal D}\bar h^{\wt\rmtr\rmd\rmT}_{\mu D}\over\sqrt{4\pi\hbar}},
&\eqnon\FCVb\cr
}$$
where $\wt v^\rmT$ is any transverse vector field
without $\ell=1$ modes. Therefore, Eqs.~\FCVa\ and \FCVb\ give
$$\eqalignno{
&\quad
  \prod_\tau
  \Big[
  \Big( {\Delta\tau a^{D-3}(\tau)\over2\pi\hbar} \Big)^{1/2}
  {\cal D}h^\rmT_{iD}(\tau)
  \Big]
  \Big[
  \Big(
  {\Delta\tau a^{D-5}(\tau)\over2\pi\hbar
  \big( -\wt\square-(D-2) \big) }
  \Big)^{1/2}
  {\cal D}w^\rmT_i(\tau)
  \Big]
\cr
&=
  \bigg\vert\det\bigg[ g^{\mu\nu}\bigg(
  { -\square-(D-1)r^{-2}\over -\square+r^{-2} }
  \bigg) \bigg](\wt v^\rmT)
  \bigg\vert^{1/2}
  {{\cal D}\bar h^{\wt\rmtr\rmd\rmT}_{\mu D}\over\sqrt{4\pi\hbar}}
  {{\cal D}\bar\xi^\rmT_\mu\over\sqrt{4\pi\hbar}}.
&\eqnon\FCVM\cr
}$$
Instead of Eqs.~\KanxiTi\ and \KanhTiD,
if Eqs.~\KanhtrdTiD\ and \KanwTi\ are used
through the following steps of the change of variables, i.e.,
$
  \big( h^\rmT_{iD},w^\rmT_i \big)
 \rightarrow
  \big( \bar h^{\wt\rmtr\rmd\rmT}_{\ \mu D},w^\rmT_i \big)
 \rightarrow
  \big( \bar h^{\wt\rmtr\rmd\rmT}_{\ \mu D},\bar\xi^\rmT_\mu \big)
$,
the different result from Eq.~\FCVM\ is derived:
$$\eqalignno{
&\quad
  \prod_\tau
  \Big[
  \Big( {\Delta\tau a^{D-3}(\tau)\over2\pi\hbar} \Big)^{1/2}
  {\cal D}h^\rmT_{iD}(\tau)
  \Big]
  \Big[
  \Big(
  {\Delta\tau a^{D-5}(\tau)\over2\pi\hbar
  \big( -\wt\square-(D-2) \big) }
  \Big)^{1/2}
  {\cal D}w^\rmT_i(\tau)
  \Big]
\cr
&=
  \bigg\vert\det\bigg[ g^{\mu\nu}\bigg(
  { -\square-(D-1)r^{-2} \over -\square+r^{-2} }
  \bigg) \bigg](\wt v^\rmT)
  \bigg\vert^{-1/2}
  {{\cal D}\bar h^{\wt\rmtr\rmd\rmT}_{\mu D}\over\sqrt{4\pi\hbar}}
  {{\cal D}\bar\xi^\rmT_\mu\over\sqrt{4\pi\hbar}}.
&\eqnon\SCVM\cr
}$$

{}From Eqs.~\FCVM\ and \SCVM, therefore, the factor
$$\eqalignno{
&
  \bigg\vert\det\bigg[ g^{\mu\nu}\bigg(
  { -\square-(D-1)r^{-2} \over -\square+r^{-2} }
  \bigg) \bigg](\wt v^\rmT)
  \bigg\vert^{1/2}
&\eqnon\AJF\cr
}$$
might be naively thought to take a trivial value 1 and both measures
for tensor fields defined in terms of transverse vector fields
without $\ell\ne1$ modes of $\wt\square$ on the unit $S^{D-1}$
would be concluded to be equal to each other.
However, the factor is obviously the same
with the uncancellation factor \UCFc\ excluding the $\ell=1$ mode part,
and in order to actually calculate the factor \AJF\
one need to regularize it because the arguments of the determinates
are infinite matrices.
Unfortunately, some regularization fail to make it 1 [\GrKo].
\vskip 0.5cm
{\bf 4.3\quad Anomalous Jacobian Factor of Scalar Part}

The relations among scalar fields
$\bar h$, $\bar\phi$, $\bar\xi$, $\bar h^{\wt\rmtr\rmd\rmL}_{DD}$,
$h$, $h^{\wt\rmtr}_{DD}$, $\Phi$ and $\psi$ are
$$\eqalignno{
&
  h
 =
  \sqrt D \bar h,
&\eqnon\hKan\cr
&
  h^{\wt\rmtr}_{DD}
 =
  {1\over D}\big( D{\partial_D}^2-\square \big)O^s_\phi\bar\phi
  +a^{-2}O^s_{\wt\rmtr\rmd\rmL}\bar h^{\wt\rmtr\rmd\rmL}_{DD}
  +2a^{-1}\big( \partial_D-a^{-1}\dot a \big)
  O^{\bar vs}_{\xi^\rmd}O^s_{\xi^\rmL}\bar\xi,
&\eqnon\KKRELhDD\cr
&
  \wt\square^{-1}\Phi
 =
  \big( \partial_D-a^{-1}\dot a \big)O^s_\phi\bar\phi
  -\wt\square^{-1}
  \big( \partial_D+(D-2)a^{-1}\dot a \big)
  O^s_{\wt\rmtr\rmd\rmL}\bar h^{\wt\rmtr\rmd\rmL}_{DD}
\cr
&\quad
  -\wt\square^{-1}
  a\big(
  \square
  -2a^{-1}\dot a\partial_D
  -2(D-2)a^{-2}\dot a^2
  -(D-2)r^{-2}
  -2a^{-2}\wt\square
  \big)
  O^{\bar vs}_{\xi^\rmd}O^s_{\xi^\rmL}\bar\xi,
&\eqnon\KKRelPhi\cr
&
  {D-1\over D-2}(\wt\square+D-1)^{-1}\psi
 =
  \wt\square O^s_\phi\bar\phi
  +{D-1\over D-2}
  (\wt\square+D-1)^{-1}a^2
  \Big[
  \square
  +(D-2)a^{-1}\dot a\partial_D
\cr
&\qquad
  +(D-2)^2a^{-2}\dot a^2
  -(D-2)r^{-2}
  -{D-2\over D-1}a^{-2}\wt\square
  \Big]
  O^s_{\wt\rmtr\rmd\rmL}\bar h^{\wt\rmtr\rmd\rmL}_{DD}
\cr
&\quad
  -2a\big( \partial_D+(D-2)a^{-1}\dot a \big)
  O^{\bar vs}_{\xi^\rmd}O^s_{\xi^\rmL}\bar\xi,
&\eqnon\KKRelPsi\cr
}$$
where
$
  O_{\xi^\rmd}^{\bar vs}
 \=
  \Big( 2\big( -\square-Dr^{-2} \big) \Big)^{-1/2}
$.

With respect to the trace part,
from the relation \hKan,
we have the following relationship of measures;
$$\eqalignno{
&\quad
  \prod_\tau
  \Big[
  \Big( {\Delta\tau(CD+2)a^{D-1}(\tau)\over8\pi\hbar D} \Big)^{1/2}
  {\cal D}h(\tau)
  \Big]
 =
  {{\cal D}\bar h\over\sqrt{8\pi\hbar/(CD+2)}},
&\eqnon\AJtrap\cr
}$$
which means that no anomalous thing happens
under the change of variables in this part.

When turning our discussion to the change of variables of
$\bar h^{\wt\rmtr\rmd\rmL}_{DD}$, $\bar \xi$, $\bar \phi$
and $h^{\wt\rmtr}_{DD}$, $\Phi$, $\psi$,
it is important to note that
$h^{\wt\rmtr}_{DD}$ and $\bar \phi$
have both modes of $\ell=0$ and 1,
and the variables $\Phi$, $\bar \xi$ have only $\ell=1$ modes
among them,
on the other hand,
$\psi$ and $\bar h^{\wt\rmtr\rmd\rmL}_{DD}$
have no one of these modes.
In this paper, we study the Jacobian factor
with respect to only the part
having no modes which are $\ell=0$ and 1 of $\wt\square$
on the unit $S^{D-1}$ in order to check the relation between it
and the uncancellation factor \UCFc:
first, from Eqs.~\KKRELhDD, \KKRelPhi\ and \KKRelPsi, we have
$$\eqalignno{
&\quad
  \Phi
  -{D-1\over D-2}(\wt\square+D-1)^{-1}
  \big( \partial_D-a^{-1}\dot a \big)
  \psi
\cr
&=
  {D-1\over D-2}(\wt\square+D-1)^{-1}a^2
  \big( \partial_D+(D-1)a^{-1}\dot a \big)
  \big( -\square+(D-2)r^{-2} \big)
  O^s_{\wt\rmtr\rmd\rmL}\bar h^{\wt\rmtr\rmd\rmL}_{DD}
\cr
&\quad
  -a\big(
  -\square
  +(D-2)r^{-2}
  \big)
  O^{\bar vs}_{\xi^\rmd}O^s_{\xi^\rmL}\bar\xi,
&\eqnon\FKEISAN\cr
&\quad
  D\wt\square
  h^{\wt\rmtr}_{DD}
  -{D-1\over D-2}(\wt\square+D-1)^{-1}
  \big( D{\partial_D}^2-\square \big)\psi
\cr
&=
  -{(D-1)^2\over D-2}a^2(\wt\square+D-1)^{-1}
  \big( -\square+2(D-1)r^{-2}
  -2a^{-1}\dot a\partial_D+2a^{-2}\wt\square
  \big)
\cr
&\qquad
  \times
  \big( -\square+(D-2)r^{-2} \big)
  O^s_{\wt\rmtr\rmd\rmL}\bar h^{\wt\rmtr\rmd\rmL}_{DD}
\cr
&\quad
  -2(D-1)a\partial_D\big( -\square+(D-2)r^{-2} \big)
  O^{\bar vs}_{\xi^\rmd}O^s_{\xi^\rmL}\bar\xi,
&\eqnon\SKEISAN\cr
}$$
and these equations give
$$\eqalignno{
&\quad
  2(D-1)a\partial_Da^{-1}\Phi
  -D\wt\square h^{\wt\rmtr}_{DD}
  -{D-1\over D-2}(\wt\square+D-1)^{-1}
\cr
&\qquad
  \times
  \Big[
  (D-2){\partial_D}^2
  -4(D-1)a^{-1}\dot a\partial_D
  +2(D-1)( a^{-2}\dot a^2+a^{-2} )
  +\square
  \Big]\psi
\cr
&=
  {(D-1)^2\over D-2}a^2(\wt\square+D-1)^{-1}
  \square
  \big( -\square+(D-2)r^{-2} \big)
  O^s_{\wt\rmtr\rmd\rmL}\bar h^{\wt\rmtr\rmd\rmL}_{DD}.
&\eqnon\KOUSHIKI
}$$
Now let us change variables, using
Eqs.~\KOUSHIKI, \FKEISAN\ and \KKRelPsi\ through the following steps:
$
  ( h^{\wt\rmtr}_{DD},\Phi,\psi )
 \rightarrow
  ( \bar h^{\wt\rmtr\rmd\rmL}_{DD},\Phi,\psi )
 \rightarrow
  ( \bar h^{\wt\rmtr\rmd\rmL}_{DD},\bar \xi,\psi )
 \rightarrow
  ( \bar h^{\wt\rmtr\rmd\rmL}_{DD},\bar \xi,\bar \phi )
$.
Then, an anomalous Jacobian factor is derived:
$$\eqalignno{
&\quad
  \prod_\tau
  \Big[
  \Big( {\Delta\tau Da^{D-1}(\tau)\over4\pi\hbar(D-1)} \Big)^{1/2}
  {\cal D}h^{\wt\rmtr}_{DD}(\tau)
  \Big]
  \Big[
  \Big( {\Delta\tau a^{D-3}(\tau)
  \over2\pi\hbar(-\wt\square) } \Big)^{1/2}
  {\cal D}\Phi(\tau)
  \Big]
\cr
&\qquad
  \times
  \Big[
  \Big(
  {\Delta\tau(D-1)a^{D-5}(\tau)\over4\pi\hbar(D-2)
  ( \wt\square+D-1 )\wt\square
  } \Big)^{1/2}
  {\cal D}\psi(\tau)
  \Big]
\cr
&=
  \bigg\vert\det\bigg[
  { -\square+(D-2)r^{-2} \over -\square-Dr^{-2} }
  \bigg](\bar s)\bigg\vert
  {{\cal D}\bar h^{\wt\rmtr\rmd\rmL}_{DD}\over\sqrt{4\pi\hbar}}
  {{\cal D}\bar\xi\over\sqrt{4\pi\hbar}}
  {{\cal D}\bar\phi\over\sqrt{4\pi\hbar}},
&\eqnon\FRAJSP\cr
}$$
where $\bar s$ is any scalar fields
without $\ell=0$ and 1 of $\wt\square$.
While, if the other steps are used together with other equations
instead of Eqs.~\KOUSHIKI, \FKEISAN\ and \KKRelPsi,
we may obtain another Jacobian factor which is
inverse of the Jacobian factor in \FRAJSP,
because, as mentioned in subsection 4.1, a naive Jacobian factor
under the change of variables
between the covariant and canonical path-integral measures
would be suggested to be 1
according to the truth that both measures are defined
by the Gaussian integral.
Thus the Jacobian factor in \FRAJSP\ may be regarded
as an anomalous one.

Furthermore, we may remember the following anomalous Jacobian factor
in path-integral measures
for the Abelian gauge theory on de~Sitter background;
$$\eqalignno{
&\quad
  \bigg\vert\det\bigg[
  { -\square \over -\square+(D-2)r^{-2} }
  \bigg](\wt s)\bigg\vert,
&\eqnon\AJGT\cr
}$$
where $\wt s$ is a scalar field
with no mode having $\ell=0$ of $\wt\square$,
thus the Jacobian factors in Eqs.~\FRAJSP\ and \AJGT\
give an anomalous factor:
$$\eqalignno{
&\quad
  \bigg\vert\det\bigg[
  { -\square \over -\square-Dr^{-2} }
  \bigg](\bar s)\bigg\vert,
&\eqnon\AJLGSP\cr
}$$
which is equal to
$$\eqalignno{
&\quad
  \bigg\vert\det\bigg[
  g^{\mu\nu}\bigg({ -\square-(D-1)r^{-2} \over -\square+r^{-2} }\bigg)
  \bigg](\bar v^\rmL)\bigg\vert,
&\eqnon\AJLGSP\cr
}$$
because the Jacobian factor can be separated
into respectively independent parts of modes of $\wt\square$.
The last equation \AJLGSP\ is nothing but
the longitudinal part of the uncancellation factor \UCFc\
excluding the $\ell=1$ modes.

\chapter{Conclusion}
In this paper, the conformal rotation in the Euclidean path integral
of linearized gravity around a saddle-point background
with the cosmological term was discussed and that
it in a nontrivial curved spacetime will be more complicated
than the flat spacetime because the sign of the mass term
in the action $I^\psi$ (see Eq.~\DecI) is same
with ordinary scalar fields, in spite of the opposite sign
of the kinetic term.
But the action of physical variable might be positive definite
so that the path integral may converge.
Furthermore, it was also argued
that the covariant path-integral measure
does not give us the physical one-loop partition function
and particularly in de~Sitter spacetime
the uncancellation factor was shown to be equal to
an anomalous Jacobian factor under the change of variables
between the covariant and canonical path-integral measures.

Although in this paper
we have discussed with respect to only tensor fields,
the Faddeev-Popov vector fields may be actually shown
similarly to have an anomalous Jacobian factor
between the covariant and canonical measures.
Furthermore, the anomalous factor is the same with
one in the case of ordinary vector fields.

Finally, we may note that
the difference between the path-integral measures
up to the anomalous Jacobian factor, as discussed in Ref.~\FuS,
calls the reexamination of the Faddeev-Popov conjecture to mind,
since the covariant decomposition is suitable one
for the covariant gauge and the canonical decomposition is
suitable for a special gauge fixing condition so that
redundant variables all may vanish,
so two path integrals defined in those gauge fixing conditions
are different from each other
owing to the anomalous Jacobian factor.
\endpage
\Appendix{A}
In this appendix we summarize
the scalar, vector and symmetric tensor eigenfunctions
of the covariant d'Alembertian (Laplacian) on spheres [\FuF].
In particular, eigenfunctions on $S^D$ of radius $r$
are expressed in terms of eigenfunctions
of $\wt\square$ on the unit $S^{D-1}$
and the Gegenbauer polynomials $C_m^n(x)$.

On $S^D$ of radius $r$,
eigenvalues of ${\square}$ acting on scalar functions are
$$\eqalignno{
&
  -r^{-2}\lambda_L(D,0)
 \=
  -r^{-2}L(L+D-1),
\qquad
  \rmfor\
  L
 =
  0,\ 1,\ 2,\ ...\ ,
&\eqnon\EivSc\cr
}$$
with the degeneracy $d_L(D,0)$ defined by
$$\eqalignno{
  d_L(D,0)
&\=
  \left( \matrix{ D +L \cr
                     L \cr}
                     \right)
  -\left( \matrix{ D +L -2 \cr
                      L -2 \cr}
                         \right),
&\eqnon\DegSc\cr
}$$
and the associated eigenfunctions $\phi^{L\ell m}$ can be expressed
in terms of eigenfunctions ${\wt S}^{\ell m}$ on the unit $S^{D-1}$
and the Gegenbauer polynomials $C_m^n(x)$ as
$$\eqalign{
  \phi^{L\ell m}
&\=
  f^{L\ell}(\tau/r){\wt S}^{\ell m},
\qquad
  \rmfor\ \
  L
 =
  0,\ 1,\ 2,\ ...\ ,
\cr
  \ell
&=
  0,\ 1,\ 2,\ ...\ ,\ L,
\quad
  \rmand\
  m
 =
  1,\ 2,\ 3,\ ...\ ,\ d_\ell(D-1,0),
}\eqn\ellmSSc$$
where
$$\eqalign{
&\hskip 1.5cm
  f^{L\ell}(\theta)
\=
  C_{L\ell}(D,0)(\sin\theta)^\ell
  C^{(D-1)/2+\ell}_{L-\ell}(\cos\theta),
\cr
&
  C_{L\ell}(D,0)
\=
  r^{-D/2}
  \bigg( {2^{2\ell+D-2}(L+{D-1\over 2})(L-\ell)!
  \over\pi(L+\ell+D-2)!}
  \bigg)^{1/2}
  \Gamma\bigg( \ell+{D-1\over 2} \bigg),
\cr
}\eqn\fLell
$$
and ${\wt S}^{\ell m}$ are scalar eigenfunctions of $\wt\square$
with the eigenvalue $-\lambda_\ell(D-1,0)$
and the degeneracy $d_\ell(D-1,0)$.
The indices $\ell$ and $m$ of $\phi^{L\ell m}$
denote the associated degeneracy, hence we have
a directly checkable relation such that
$$
  d_L(D,0)
  =
  \sum_{\ell=0}^Ld_\ell(D-1,0).
\eqn\DegScSSc
$$
{}From \fLell, $f^{L\ell}$ are found to satisfy
$$\eqalignno{
&
  a^{-(D-1)}{\rmd\over\rmd\tau}
  (a^{D-1}{\rmd\over\rmd\tau}f^{L\ell})
  -\lambda_\ell(D-1,0)a^{-2}f^{L\ell}
 =
  -r^{-2}\lambda_L(D,0)f^{L\ell},
&\eqnon\EiEqfLell\cr
}$$
which is an expression of eigenvalue equations obeyed
by $\phi^{L\ell m}$ in terms of $f^{L\ell}$,
and they also satisfy
$$\eqalignno{
&
  \int_0^{r\pi}\rmd\tau a^{D-1}
  f^{L\ell}f^{L'\ell}
 =
  \delta_{LL'}.
&\eqnon\COfLell\cr
}$$
{}From this and
orthogonality of ${\wt S}^{\ell m}$ on $S^{D-1}$,
the scalar eigenfunctions $\phi^{L\ell m}$ form
the orthonormal basis of scalar fields on $S^D$ of radius $r$.

Any vector field on $S^D$ can be decomposed
into the gradient of a scalar and a divergenceless vector.
Therefore the orthonormal basis of vector fields on $S^D$
is made of these two parts:
first, gradients of scalar eigenfunctions, i.e.,
$$\eqalign{
  T^{(\rms)L\ell m}_{\quad \mu}
&\=
  \big( r^{-2}\lambda_L(D,0) \big)^{-1/2}
  \nabla_\mu\phi^{L\ell m},
\qquad
  \rmfor\
  L
 =
  1,\ 2,\ 3,\ ...\ ,
\cr
  \ell
&=
  0,\ 1,\ 2,\ ...\ ,\ L,
\quad
  \rmand\
  m
 =
  1,\ 2,\ 3,\ ...\ ,\ d_\ell(D-1,0),
}\eqn\ellmScVec$$
satisfy the eigenvalue equation
$$\eqalign{
  {\square}T^{(\rms)L\ell m}_{\quad \mu}
&=
  -r^{-2}\lambda_L(D,1\rms)T^{(\rms)L\ell m}_{\quad \mu},
\cr
  \lambda_L(D,1\rms)
&\=
  L(L+D-1)-(D-1),
\cr
}\eqn\EivScVec
$$
and the orthogonality.
Note here that there exists no $T^{(\rms)L\ell m}_{\quad \mu}$
for $L=0$ because it trivially vanishes
and also that the eigenvalues $\lambda_L(D,1\rms)$
are thus positive for $L\ge1$.
Degeneracies of $\square$ acting on gradients of scalars on $S^D$
are $d_L(D,1\rms)\=d_L(D,0)$ for $L\neq0$.
Next, eigenvalues of $\square$ acting
on divergenceless vectors on $S^D$ of radius $r$ are
$$\eqalignno{
  -r^{-2}\lambda_L(D,1\rmd)
&\=
  -r^{-2}\big( L(L +D -1) -1 \big),
\qquad
  \rmfor\
  L
 =
  1,\ 2,\ 3,\ ...\ ,
&\eqnon\EivDivVec\cr
}$$
with the degeneracy $d_L(D,1\rmd)$ defined by
$$\eqalignno{
  d_L(D,1\rmd)
&\=
  (D+1)d_L(D,0)-d_{L-1}(D,0)-d_{L+1}(D,0).
&\eqnon\Degdv\cr
}$$
Furthermore, any divergence vector can be split
into the longitudinal
and transverse parts,
thus the divergenceless vector eigenfunctions
$T^{(\rmd)L\ell m}_{\quad \mu}$
consist of their parts:
the longitudinal vector eigenfunctions
$T^{(\rmL)L\ell m}_{\quad \mu}$ are defined
by scalar eigenfunctions on $S^D$;
$$\eqalign{
  T^{(\rmL)L\ell m}_{\quad D}
&\=
  C_{L\ell}(D,1\rmL)a^{-1}\phi^{L\ell m},
\cr
  T^{(\rmL)L\ell m}_{\quad i}
&\=
  -{\wt\nabla}_i{\wt\square}^{-1}
  a^{-(D-3)}\partial_D(a^{D-1}T^{(\rmL)L\ell m}_{\quad D}),
\qquad
  \rmfor\
  L
 =
  1,\ 2,\ 3,\ ...\ ,
\cr
  \ell
&=
  1,\ 2,\ 3,\ ...\ ,\ L,
\quad
  \rmand\
  m
 =
  1,\ 2,\ 3,\ ...\ ,\ d_\ell(D-1,0),
}\eqn\ellmLonVec$$
with the normalization constants
$$
  C_{L\ell}(D,1\rmL)
 =
  \Big( r^2\ell(\ell+D-2)[L(L+D-1)+D-2]^{-1}
  \Big)^{1/2},
\eqn\CLellLon
$$
where we note that there exists no $T^{(\rmL)L\ell m}_{\quad \mu}$
for $\ell=0$ because the operator $\wt\square^{-1}$
or the coefficient $C_{L\ell}(D,1\rmL)$ does not permit it.
This means that the degeneracy of
the longitudinal vector eigenfunctions becomes
$$\eqalignno{
  d_L(D,1\rmL)
&=
  \sum_{\ell=1}^Ld_\ell(D-1,0)
 =
  d_L(D,0) -1.
&\eqnon\DegLonVec\cr
}$$
As for the transverse vector eigenfunctions
$T^{(\rmT)L\ell m}_{\quad \mu}$,
they are expressed in terms of
divergenceless vector eigenfunctions $\wt S^{(\rmd)\ell m}_{\quad i}$
of $\wt\square$ on the unit $S^{D-1}$ with the eigenvalue
$-\lambda_\ell(D-1,1\rmd)$ and the degeneracy $d_\ell(D-1,1\rmd)$,
and $f^{L\ell}$ as
$$\eqalign{
  T^{(\rmT)L\ell m}_{\quad D}
&\=
   0,
\qquad
  T^{(\rmT)L\ell m}_{\quad i}
 \=
  af^{L\ell}\wt S^{(\rmd)\ell m}_{\quad i},
\qquad
  \rmfor\
  L
 =
  1,\ 2,\ 3,\ ...\ ,
\cr
  \ell
&=
  1,\ 2,\ 3,\ ...\ ,\ L,
\quad
  \rmand\
  m
 =
  1,\ 2,\ 3,\ ...\ ,\ d_\ell(D-1,1\rmd),
}\eqn\ellmTraVec$$
where the indices $\ell$ and $m$ of
$T^{(\rmT)L\ell m}_{\quad \mu}$ stand for
the degeneracy $d_L(D,1\rmT)$, which becomes
$$\eqalignno{
  d_L(D,1\rmT)
&=
  \sum_{\ell=1}^Ld_\ell(D-1,1\rmd)
 =
  Dd_L(D,0) -d_{L-1}(D,0) -d_{L+1}(D,0) +1.
&\eqnon\DegTraVec\cr
}$$
Both eigenfunctions
$T^{(\rmL)L\ell m}_{\quad \mu}$ and
$T^{(\rmT)L\ell m}_{\quad \mu}$
obey the eigenvalue equation that
divergenceless vector eigenfunctions satisfy.
One obtains the relation
$$\eqalignno{
  d_L(D,1\rmd)
&=
  d_L(D,1\rmL) +d_L(D,\rm1T).
&\eqnon\DegDivLonTra\cr
}$$

Any symmetric tensor field on $S^D$ can be covariantly decomposed
by Eq.~\DefhS,
thus symmetric tensor eigenfunctions of $\square$ can be constructed,
by separating them into their components, as follows:
First, eigenfunctions forming the trace component
$h^{\rmtr}_{\mu\nu}$ of tensor fields are defined by
$$\eqalign{
  T^{(\rmtr)L\ell m}_{\quad \mu\nu}
&\=
  C_L(D,2\rmtr)
  g_{\mu\nu}\phi^{L\ell m},
\qquad
  \rmfor\
  L
 =
  0,\ 1,\ 2,\ ...\ ,
\cr
  \ell
&=
  0,\ 1,\ 2,\ ...\ ,\ L,
\quad
  \rmand\
  m
 =
  1,\ 2,\ 3,\ ...\ ,\ d_\ell(D-1,0),
\cr
&
  \rmwith\
  C_L(D,2\rmtr)
 \=
  {1\over\sqrt D},
\cr
}\eqn\DefYtr
$$
their eigenvalues and degeneracies for
the d'Alembertian $\square$ obviously become
$$\eqalign{
  -r^{-2}\lambda_L(D,2\rmtr)
&\=
  -r^{-2}L(L +D -1),
\qquad
  d_L(D,2\rmtr)
 \=
  d_L(D,0).
\cr
}\eqn\Degttr
$$
Secondly, the component $h^{\wt\rmtr\phi}_{\mu\nu}$
defined in Eq.~\Defhwtrp\ gives the following eigenfunctions;
$$\eqalign{
  T^{(\wt\rmtr\rms)L\ell m}_{\quad \mu\nu}
&\=
  C_L(D,2\wt\rmtr\rms)
  \Big(
  \nabla_\mu\nabla_\nu
  -{1\over D}g_{\mu\nu}\square
  \Big)
  \phi^{L\ell m},
\qquad
  \rmfor\
  L
 =
  2,\ 3,\ 4,\ ...\ ,
\cr
  \ell
&=
  0,\ 1,\ 2,\ ...\ ,\ L,
\quad
  \rmand\
  m
 =
  1,\ 2,\ 3,\ ...\ ,\ d_\ell(D-1,0),
\cr
&\hskip -1.0cm
  \rmwith\
  C_L(D,2\wt\rmtr\rms)
 \=
  \bigg(
  {r^4D\over(D-1)
  L(L+D-1)\big( L(L+D-1)-D \big)}
  \bigg)^{1/2},
}\eqn\DefYwtrs
$$
and the eigenvalues and degeneracies, respectively, are
$$\eqalign{
  -r^{-2}\lambda_L(D,2\wt\rmtr\rms)
&\=
  -r^{-2}\big( L(L +D -1)-2D \big),
\qquad
  d_L(D,2\wt\rmtr\rms)
 \=
  d_L(D,0).
\cr
}\eqn\Degttrls
$$
Thirdly, the eigenfunctions
$T^{(\wt\rmtr v^\rmd)L\ell m}_{\quad \mu\nu}$
forming the component $h^{\wt\rmtr\xi^\rmd}_{\mu\nu}$ of tensor fields
obey the eigenvalue equation
$$\eqalign{
  \square
  T^{(\wt\rmtr v^\rmd)L\ell m}_{\quad \mu\nu}
&=
  -r^{-2}\lambda_L(D,2\wt\rmtr v^\rmd)
  T^{(\wt\rmtr v^\rmd)L\ell m}_{\quad \mu\nu},
\cr
  -r^{-2}\lambda_L(D,2\wt\rmtr v^\rmd)
&\=
  -r^{-2}\big( L(L+D-1)-(D+2) \big),
\cr
}\eqn\EIEQtrlvdT
$$
with the degeneracy
$$\eqalign{
  d_L(D,2\wt\rmtr v^\rmd)
&\=
  d_L(D,1\rmd),
\qquad
  \rmfor\
  L
 =
  2,\ 3,\ 4,\ ...\ ,
}\eqn\Eigtrldv
$$
and they are given, from Eq.~\DefhwtrxdIJ, by
$$\eqalign{
&
  T^{(\wt\rmtr v^\rmd)L\ell m}_{\quad \mu\nu}
 \=
  C_L(D,2\wt\rmtr v^\rmd)
  \Big(
  \nabla_\mu T^{(\rmd)L\ell m}_{\quad \nu}
  +\nabla_\nu T^{(\rmd)L\ell m}_{\quad \mu}
  \Big),
\cr
&\quad
  \rmwith\
  C_L(D,2\wt\rmtr v^\rmd)
 \=
  \Big( 2r^{-2}\big( L(L+D-1)-D \big) \Big)^{-1/2},
\cr
}\eqn\DefYwtrdv
$$
which may be separated into two parts,
as mentioned in Eq.~\DefphwtrxdIJ;
$$\eqalign{
  T^{(\wt\rmtr v^\rmd)L\ell m}_{\quad \mu\nu}
&\=
  \Big(
  T^{(\wt\rmtr v^\rmL)L\ell m}_{\quad \mu\nu},
  T^{(\wt\rmtr v^\rmT)L\ell m}_{\quad \mu\nu}
  \Big),
\cr
}\eqn\DefYd
$$
where each part corresponds to
the case of that $T^{(\rmd)L\ell m}_{\quad \mu}$ in \DefYwtrdv\
is $T^{(\rmL)L\ell m}_{\quad \mu}$ or $T^{(\rmT)L\ell m}_{\quad \mu}$
and their degeneracies, respectively, are
$d_L(D,2\wt\rmtr v^\rmL)\=d_L(D,1\rmL)$ and
$d_L(D,2\wt\rmtr v^\rmT)\=d_L(D,1\rmT)$.
Thus we have the same relation as \DegDivLonTra;
$$\eqalignno{
  d_L(D,2\wt\rmtr v^\rmd)
&=
  d_L(D,2\wt\rmtr v^\rmL)+d_L(D,2\wt\rmtr v^\rmT).
&\eqnon\DegtrldT\cr
}$$
Lastly, the traceless and divergenceless symmetric tensor
eigenfunctions $T^{(\wt\rmtr\rmd)L\ell m}_{\quad \mu\nu}$
corresponding to $h^{\wt\rmtr\rmd}_{\mu\nu}$ in Eq.~\Decxi\ obey
$$\eqalign{
  {\square}T^{(\wt\rmtr\rmd)L\ell m}_{\quad \mu\nu}
&=
  -r^{-2}\lambda_L(D,2\wt\rmtr\rmd)
  T^{(\wt\rmtr\rmd)L\ell m}_{\quad \mu\nu},
\cr
  \lambda_L(D,2\wt\rmtr\rmd)
&\=
  L(L+D-1)-2,
\cr
}\eqn\DefYwtd
$$
with the degeneracy
$$\eqalign{
&
  d_L(D,2\wt\rmtr\rmd)
 =
  {(D+1)(D+2)\over2}d_L(D,0)
  -d_{L-1}(D,1\rmd)-d_{L+1}(D,1\rmd)
\cr
&\quad
  -d_{L-2}(D,0)-2d_{L}(D,0)-d_{L+2}(D,0),
\qquad
  \rmfor\
  L
 =
  2,\ 3,\ 4,\ ...\ .
\cr
}\eqn\DegtrldT
$$
Furthermore, they can be composed of three parts according to
Eqs.~\hwtrd$\sim$\htT:
$$\eqalignno{
&
  T^{(\wt\rmtr\rmd)L\ell m}_{\quad \mu\nu}
 \=
  \Big(
  T^{(\wt\rmtr\rmd\rmL)L\ell m}_{\quad \mu\nu},
  T^{(\wt\rmtr\rmd\rmT)L\ell m}_{\quad \mu\nu},
  T^{(\rmt\rmT)L\ell m}_{\quad \mu\nu}
  \Big),
&\eqnon\DefTtrld\cr
}$$
where each part is defined by
$$\eqalign{
  T^{(\wt\rmtr\rmd\rmL)L\ell m}_{\quad DD}
&\=
  C_{L\ell}(D,2\wt\rmtr\rmd\rmL)
  a^{-2}\phi^{L\ell m},
\cr
  T^{(\wt\rmtr\rmd\rmL)L\ell m}_{\quad iD}
&\=
  -\wt\nabla_i\wt\square^{-1}
  a^{-(D-2)}\partial_D
  \big( a^DT^{(\wt\rmtr\rmd\rmL)L\ell m}_{\quad DD} \big)
\cr
  T^{(\wt\rmtr\rmd\rmL)L\ell m}_{\quad ij}
&\=
  {D-1\over D-2}
  \Big(
  \wt\nabla_i\wt\nabla_j\wt\square^{-1}-{1\over D-1}\wt g_{ij}
  \Big)
  ( \wt\square+D-1 )^{-1}
\cr
&\hskip -1.5cm
  \times
  \Big[
  a^{-(D-3)}\partial_D
  \big( a\partial_D( a^DT^{(\wt\rmtr\rmd\rmL)L\ell m}_{\quad DD} )\big)
  +{a^2\over D-1}\wt\square T^{(\wt\rmtr\rmd\rmL)L\ell m}_{\quad DD}
  \Big]
  -{a^2\over D-1}\wt g_{ij}T^{(\wt\rmtr\rmd\rmL)L\ell m}_{\quad DD},
\cr
  \rmfor\
  \ell
&=
  2,\ 3,\ 4,\ ...,\ L,
\quad
  \rmand\
  m
 =
  1,\ 2,\ 3,\ ...,\ d_\ell(D-1,0),
\cr
&\hskip -1.0cm
  \rmwith\
  C_{L\ell}(D,2\wt\rmtr\rmd\rmL)
 =
  \bigg[
  {
  r^4(D-2)\lambda_\ell(D-1,0)\big( \lambda_\ell(D-1,0)-(D-1) \big)
  \over
  (D-1)\lambda_L(D,0)\big( \lambda_L(D,0)+(D-2) \big)
  }
  \bigg]^{1/2},
\cr
}\eqn\DefYtTLM
$$
$$\eqalign{
  T^{(\wt\rmtr\rmd\rmT)L\ell m}_{\quad \mu D}
&\=
  C_{L\ell}(D,2\wt\rmtr\rmd\rmT)
  a^{-1}
  T^{(\rmT)L\ell m}_{\quad \mu},
\cr
  T^{(\wt\rmtr\rmd\rmT)L\ell m}_{\quad ij}
&\=
  -\wt\nabla_i( \wt\square+D-2 )^{-1}
  a^{-(D-3)}\partial_D
  ( a^{D-1}T^{(\wt\rmtr\rmd\rmT)L\ell m}_{\quad jD} )
\cr
&\quad
  -\wt\nabla_j( \wt\square+D-2 )^{-1}
  a^{-(D-3)}\partial_D
  ( a^{D-1}T^{(\wt\rmtr\rmd\rmT)L\ell m}_{\quad iD} ),
\cr
  \rmfor\
  \ell
&=
  2,\ 3,\ 4,\ ...,\ L,
\quad
  \rmand\
  m
 =
  1,\ 2,\ 3,\ ...,\ d_\ell(D-1,1\rmd),
\cr
&\hskip -1.0cm
  \rmwith\
  C_{L\ell}(D,2\wt\rmtr\rmd\rmT)
 =
  \bigg[
  {
  r^2\big( \lambda_\ell(D-1,1\rmd)-(D-2) \big)
  \over
  2\big( \lambda_L(D,1\rmT)+1 \big)
  }
\bigg]^{1/2},
\cr
}\eqn\DefYTDT
$$
and
$$\eqalign{
  T^{(\rmt\rmT)L\ell m}_{\quad \mu D}
&\=
  0,
\qquad
  T^{(\rmt\rmT)L\ell m}_{\quad ij}
 \=
  a^2f^{L\ell}
  \wt S^{(\wt\rmtr\rmd)\ell m}_{\quad ij},
\cr
  \rmfor\
  \ell
&=
  2,\ 3,\ 4,\ ...,\ L,
\quad
  \rmand\
  m
 =
  1,\ 2,\ 3,\ ...,\ d_\ell(D-1,2\wt\rmtr\rmd),
\cr
&
}\eqn\DefTtT
$$
where $\wt S^{(\wt\rmtr\rmd)\ell m}_{\quad ij}$
are traceless and divergenceless symmetric tensor eigenfunctions
of $\wt\square$
on the unit $S^{D-1}$.
The degeneracies of their parts are
$$\eqalignno{
&
  d_L(D,2\wt\rmtr\rmd\rmL)
 \=
  \sum_{\ell=2}^{L}
  d_\ell(D-1,0)
 =
  d_L(D,0)
  -(D+1),
&\eqnon\DegtrldL\cr
&
  d_L(D,2\wt\rmtr\rmd\rmT)
 \=
  \sum_{\ell=2}^{L}
  d_\ell(D-1,1\rmd)
\cr
&\quad
 =
  Dd_L(D,0)
  -d_{L-1}(D,1\rmL)
  -d_{L+1}(D,1\rmL)
  -{D^2-D+2\over2},
&\eqnon\DegtrldT\cr
&
  d_L(D,2\rmt\rmT)
 \=
  \sum_{\ell=2}^{L}
  d_\ell(D-1,2\wt\rmtr\rmd)
\cr
&\quad
 =
  {D(D+1)\over2}
  d_L(D,0)
  -d_{L-1}(D,1\rmT)
  -d_{L+1}(D,1\rmT)
  -d_{L-2}(D,0)
\cr
&\qquad
  -2d_L(D,0)
  -d_{L+2}(D,0)
  +(D+1)
  +{D^2-D+2\over2},
&\eqnon\DegtT\cr
}$$
and it is easily found to satisfy
$$\eqalignno{
  d_L(D,2\wt\rmtr\rmd)
&=
  d_L(D,2\wt\rmtr\rmL)
  +d_L(D,2\wt\rmtr\rmT)
  +d_L(D,2\rmt\rmT).
&\eqnon\Degtrld\cr
}$$

Finally, we note that the following equations identically hold:
$$\eqalign{
&
  \nabla_\mu\phi^{0\ell m}
 =
  \nabla_\mu\nabla_\nu\phi^{1\ell m}
  +r^{-2}g_{\mu\nu}\phi^{1\ell m}
 =
  \nabla_\mu T^{(\rmd)1\ell m}_{\quad \nu}
  +\nabla_\nu T^{(\rmd)1\ell m}_{\quad \mu}
 =
  0,
\cr
&
  \wt\nabla_i\wt S^{0m}
 =
  \wt\nabla_i\wt\nabla_j\wt S^{1m}
  +\wt g_{ij}\wt S^{1m}
 =
  \wt\nabla_i\wt S^{(\rmd)1m}_{\quad j}
  +\wt\nabla_j\wt S^{(\rmd)1m}_{\quad i}
 =
  0.
\cr
}\eqn\BCCond
$$
\vfill
\vfill
\vfill
\endpage

\refout
\bye